\documentclass[prl,aps,twocolumn,superscriptaddress]{revtex4-2}
\usepackage{bm}
\usepackage[colorlinks=true,linkcolor=blue,citecolor=blue]{hyperref}
\usepackage{times}
\usepackage{amsmath}
\usepackage{amssymb}
\usepackage{amsthm}
\usepackage{amsfonts}
\usepackage{enumerate}
\usepackage{latexsym}
\usepackage{ifpdf}
\newcommand{\beq}{\begin{equation}}
\newcommand{\eeq}{\end{equation}}
\usepackage{graphicx}
\usepackage{makeidx}
\usepackage{color}

\usepackage[version=4]{mhchem}
\usepackage{textcomp,mathcomp}

\begin{document}

\title{Coupled phase transitions in crystalline solids with extreme chemical disorder}

\author {Subha Dey}
\altaffiliation{Contributed equally}
\affiliation  {Department of Physics, Indian Institute of Science, Bengaluru 560012, India}
\author {Rukma Nevgi}
\altaffiliation{Contributed equally}
\affiliation  {Department of Physics, Indian Institute of Science, Bengaluru  560012, India}
\author {Suresh Chandra Joshi}
\affiliation  {Department of Physics, Indian Institute of Science, Bengaluru  560012, India}
\author {Sourav Chowdhury}
 \affiliation{Deutsches Elektronen-Synchrotron DESY, 22607 Hamburg, Germany}
\author {Nandana Bhattacharya}
\affiliation  {Department of Physics, Indian Institute of Science, Bengaluru  560012, India}
\author {Kashish Kapoor}
\affiliation  {Department of Physics, Indian Institute of Science, Bengaluru  560012, India}
\author {Tinku Dan}
\affiliation  {Deutsches Elektronen-Synchrotron DESY, 22607 Hamburg, Germany}
\author{Subhadip Chowdhury}
\affiliation{New Chemistry Unit, Jawaharlal Nehru Centre for Advanced Scientific Research, Bengaluru 560064, India}
\author{Sabyasachi Karmakar}
\affiliation{New Chemistry Unit, Jawaharlal Nehru Centre for Advanced Scientific Research, Bengaluru 560064, India}
\author{S. D. Kaushik}
\affiliation {UGC-DAE Consortium for Scientific Research Mumbai Centre, R5 Shed, Bhabha Atomic Research Centre, Mumbai 400085, India}
\author{Shibabrata Nandi}
\affiliation{Forschungszentrum Jülich GmbH, Jülich Centre for Neutron Science (JCNS-2), 52425 Jülich, Germany}
\affiliation{RWTH Aachen, Lehrstuhl für Experimentalphysik IVc, Jülich-Aachen Research Alliance (JARA-FIT), 52074 Aachen, Germany}
\author{Christoph Klewe}
\affiliation{Advanced Light Source, Lawrence Berkeley National Laboratory, Berkeley, California 94720, USA}
\author{Manuel Valvidares}
\affiliation{ALBA Synchrotron Light Source, Cerdanyola del Valles, Barcelona E-08290, Spain}
\author{Moritz Hoesch}
\affiliation{Deutsches Elektronen-Synchrotron DESY, 22607 Hamburg, Germany}
\author{George E. Sterbinsky}
\affiliation {Advanced Photon Source, Argonne National Laboratory, Lemont, Illinois 60439, USA}
\author {Srimanta Middey}
\email{smiddey@iisc.ac.in}
\affiliation {Department of Physics, Indian Institute of Science, Bengaluru 560012, India}
 
\begin{abstract}

Structural phase transitions often couple to magnetic and electronic degrees of freedom, enabling emergent phenomena in solids. In high-entropy oxides (HEOs), which typically stabilize in highly symmetric cubic phases, such transitions are considered rare due to the extreme chemical disorder-analogous to the behavior observed in high-entropy alloys. This raises a fundamental question: can the rich physics of coupled phase transitions persist in such disordered systems?  
Here, we show that targeted  design of compositionally complex oxides (CCOs) can trigger symmetry-lowering transitions, with spinel-type materials serving as a representative case.
For instance, [Mn$_{0.2}$Co$_{0.2}$Ni$_{0.2}$Cu$_{0.2}$Zn$_{0.2}$]Cr$_2$O$_4$, having two Jahn-Teller (J-T) active ions, undergoes two successive coupled structural transitions upon cooling: an orbital-driven transition at 100 K and a magnetism-driven transition at 40 K. Systematic substitution of $A$-site cations reveals that both Ni and Cu are essential for these transitions. Element specific local structure investigations uncover distinct and opposing local distortions around Ni and Cu, while Mn, Co, and Zn remain largely undistorted. These results establish that CCOs can host coupled phase transitions through ‘cooperation via competition’ among local distortions in a chemically disordered lattice. This discovery expands the design principles for complex oxides, introducing a new paradigm for tuning structural and functional properties in high-entropy systems beyond conventional symmetry constraints.

\end{abstract}

\maketitle

\begin{figure*}
\begin{center}
\includegraphics[width=0.8\textwidth ]{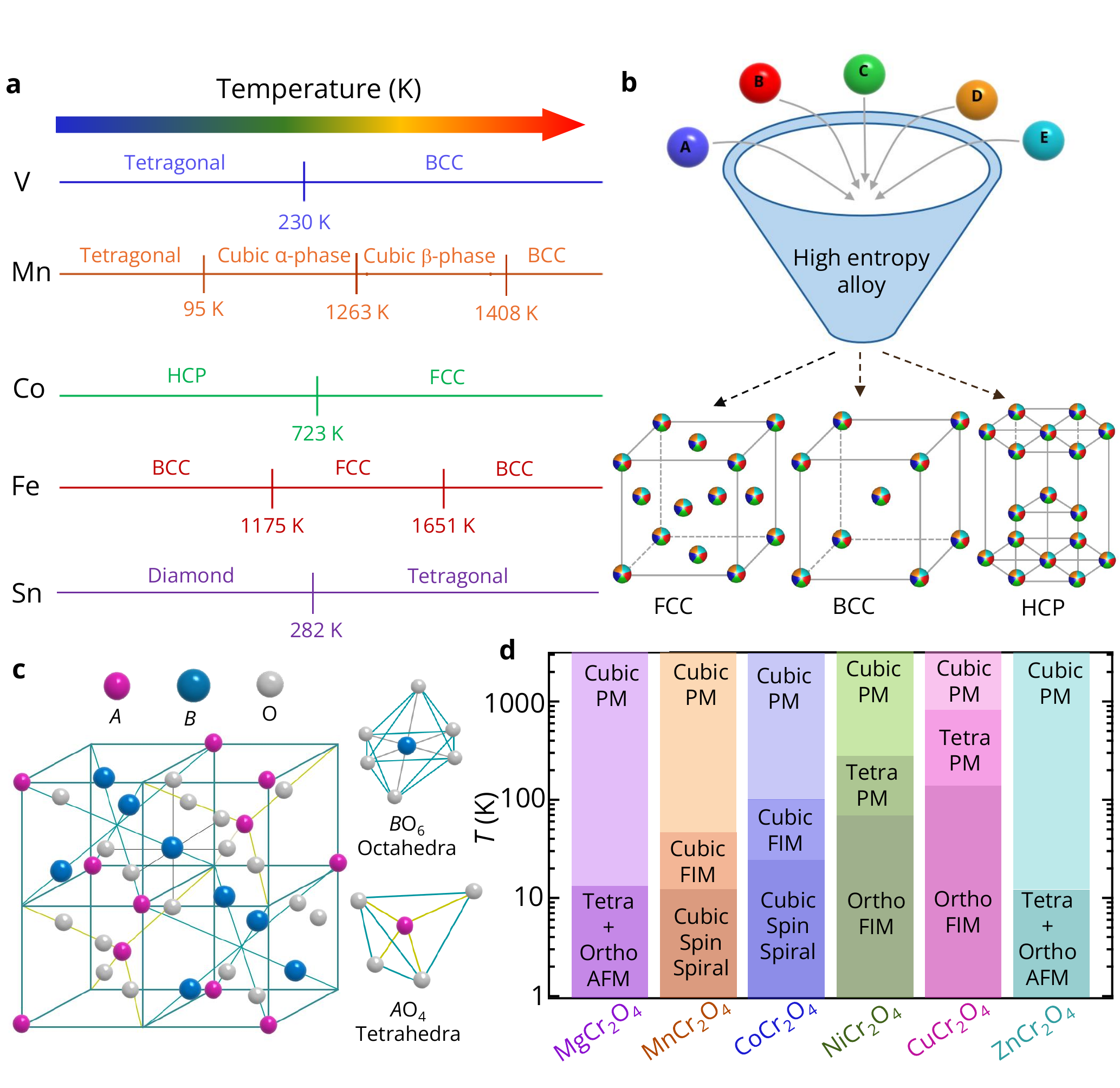}
\caption{\textbf{Structural transitions in elemental solids and chromate spinels.} \textbf{a}, Representative phase transitions in elemental solids as a function of temperature at ambient pressure.~\cite{Finkel:1969p1065,Hobbs:2003p014407,Sewak:2022p10054,Magomedov:2021p215,Sung-Ho:2010p494}  
\textbf{b}, High entropy alloys, containing five elements A, B, C, D, E, generally adopt either FCC, BCC, or HCP structures and do not show structural phase transition upon cooling.   
\textbf{c}, Crystal structure of a normal spinel $AB_2$O$_4$, showing the tetrahedral \textit{A}-site and octahedral \textit{B}-site coordination environments. \textbf{d}, Structural and magnetic phase transitions in individual \textit{A}Cr$_2$O$_4$ spinels (\textit{A} = Mg, Mn, Co, Ni, Cu, Zn), illustrating a progression from cubic paramagnetic (PM) to symmetry-lowered and magnetically ordered phases, including tetragonal and orthorhombic structures and ferri/antiferromagnetic (FIM/AFM) states.}
\label{Fig1}
\end{center}
\end{figure*}

 Structural transitions in elemental solids are ubiquitous in nature, typically occurring as a function of temperature [{\bf Fig.}\ref{Fig1}a] and pressure, which sometimes is also coupled to magnetic and electronic transition~\cite{Soderlind:1995p524, Heathman:2005p110}. In stark contrast, such structural transformations are notably rare in high entropy alloys (HEAs), which comprise five or more elements in equiatomic  or near-equiatomic  ratios~\cite{Cantor:2004p213, Yeh:2004p299, George:2019p515}. HEAs typically stabilize in high-symmetry phases such as body-centered cubic (BCC), face-centered cubic (FCC), or hexagonal close-packed (HCP) structures [{\bf Fig.}\ref{Fig1}b] and offer superior mechanical performances. In these systems, the metallic bonding combined with entropic stabilization from crystallographic disorder suppresses the thermodynamic driving force, thereby preventing structural transitions~\cite{Bhowal:2023p53}.

A similar rationale has been extended to the recently intensively studied high entropy oxides (HEOs), where temperature-driven structural transitions have been considered improbable and remain largely unreported~\cite{Rost:2015p8485,Schweidler:2024p1,Shahbazi:2025p950,Oses:2020p295,Aamlid:2023p5991}. Unlike elemental metals and alloys, where phase transitions are primarily governed by atomic arrangements and lattice strain~\cite{Christian:1956p419,Fu:2022p290}, complex oxides exhibit a richer phase landscape due to strong coupling among charge, spin, orbital, and lattice degrees of freedom~\cite{Khomskii:2014book}. These couplings can give rise to intricate, often sequential or simultaneous phase transitions involving several order parameters~\cite{Imada:1998p1039,Tokura:2017p1056, Keimer:2015p179,Post:2018p1056,Ahn:2021p1462,middey:2016p305}.
In the case of compositionally complex oxides (CCOs), the absence of global symmetry-breaking structural transitions as a function of thermodynamic parameters like temperature, pressure, etc. is attributed to competing local distortions within the lattice. These distortions arise from the individual preferences of cations-such as elongation, compression, rotation, or breathing-mode distortions of the oxygen network~\cite{Aamlid:2023p5991}. For instance, one TM ion may favor compression, while a neighboring TM ion prefers expansion, creating a form of structural “frustration” in the system. Such locally varying distortions are most effectively accommodated in a high-symmetry cubic framework.  Consequently, global structural transitions in these systems are widely considered energetically prohibitive~\cite{Kotsonis:2023p5587,Aamlid:2023p5991}. In this study, we challenge this prevailing view by demonstrating the occurrence of such transitions across a series of compounds, thereby revealing a broader property landscape than previously anticipated.

\begin{figure*}
\begin{center}
\includegraphics[width=0.8\textwidth ]{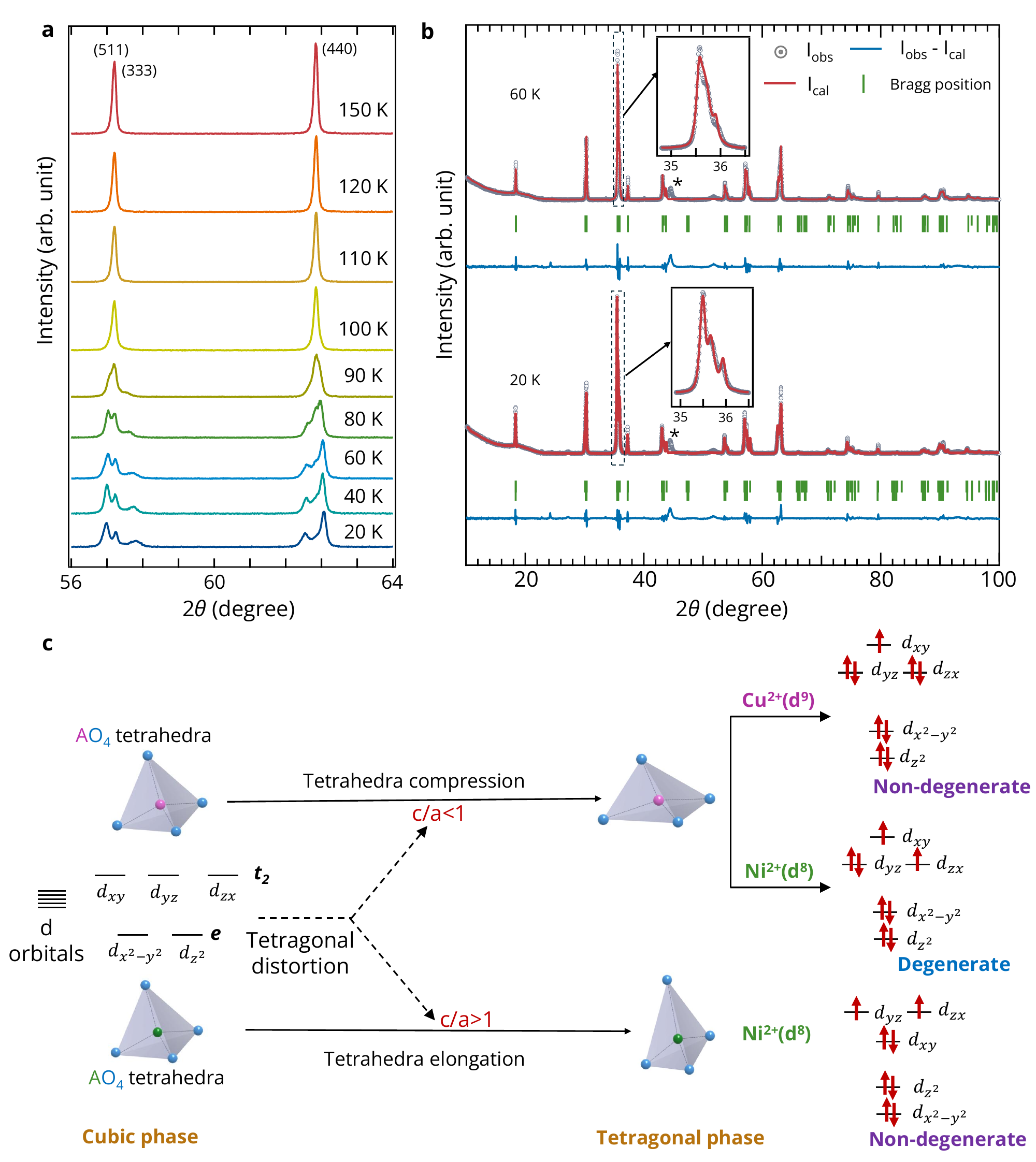}
\caption{\textbf{Structural transitions and orbital ordering.} \textbf{a}, The evolution of the selected Bragg peaks upon lowering the temperature. The splitting of the Bragg peak below 100 K indicates a structural phase transition. The fitting of the Bragg peak around 2$\theta \sim$ 57$^\circ$ requires three peaks at 60 K and four peaks at 20 K.
 \textbf{b}, XRD refinement at 60 K (cubic + tetragonal) and 20 K (tetragonal + orthorhombic). The inset shows a zoomed view of the XRD refinement of the peak with the highest intensity. * denotes the aluminum sample holder XRD peak.
 \textbf{c} Schematic of two different types of tetragonal distortion in spinel oxides. $d^9$ system favors tetrahedra compression, but $d^8$ system favors tetrahedra elongation to lift their degeneracy. The energy gap between orbital levels is not in scale.}
\label{fig: Fig 2}
\end{center}
\end{figure*}

In this work, we present direct evidence of low-temperature symmetry-lowering structural transitions through investigations of a series of compositionally complex chromate spinel oxides under a high configurational entropy setting. The parent chromate spinels $A$Cr$_2$O$_4$  exhibit diverse structural and magnetic transitions depending on the nature of the \textit{A}-site cation [{\bf Fig.}\ref{Fig1}c, d]: non-magnetic cations (e.g. $A$= Zn, Mg, Cd) induce spin-Peierls transition; magnetic but non-Jahn-Teller (J-T) active cations (e.g. $A$= Mn, Co) favor ferrimagnetism without structural change; and J-T active cations (e.g. $A$= Ni, Cu) drive sequential cubic-to-tetragonal and tetragonal-to-orthorhombic transitions~\cite{Lee:2002p1476, Bordacs:2009p077205,Jaubert:2025p086702,Suchomel:2012p054406}.
We show that a carefully engineered equiatomic combination of five distinct cations at the \textit{A}-site can drive structural symmetry-lowering transitions-but crucially, only when two J-T active ions (Ni$^{2+}$ and Cu$^{2+}$) are simultaneously present. Moreover, magnetic measurements, along with heat capacity, demonstrate the appearance of a long-range magnetic ordering, which is coupled to the appearance of the orthorhombic phase.  Compositions containing just one J-T active cation remain locked in the high-symmetry cubic phase down to the lowest measured temperatures, highlighting the essential role of cooperative orbital distortions in enabling these transitions. Temperature dependent extended X-ray absorption fine structure (EXAFS) measurements on Mn$_{0.2}$Co$_{0.2}$Ni$_{0.2}$Cu$_{0.2}$Zn$_{0.2}$Cr$_2$O$_4$  reveal distinct local environments at the \textit{A}-site: while Mn, Co, and Zn remain relatively unchanged, Cu and Ni show significant temperature-dependent opposing local distortions across the global structural change. Specifically, Cu-Cr and Cu-O bond lengths increase, while Ni-O and Ni-Cr distances decrease with cooling, identifying these two cations as the primary drivers of the observed symmetry-lowering transitions.  Our discovery of coupled phase transitions offers critical new perspectives about the underlying complexity, which will be crucial for designing CCOs with targeted properties.

{ \bf \color{blue}  Sample synthesis and characterizations:}
We synthesized polycrystalline samples of six compositionally complex spinel oxides: [Mn$_{0.2}$Co$_{0.2}$Ni$_{0.2}$Cu$_{0.2}$Zn$_{0.2}]$Cr$_2$O$_4$ (MCNCZCr$_2$O$_4$),  [Mg$_{0.2}$Co$_{0.2}$Ni$_{0.2}$Cu$_{0.2}$Zn$_{0.2}]$Cr$_2$O$_4$ (MgCNCZCr$_2$O$_4$),  [Mn$_{0.2}$Mg$_{0.2}$Ni$_{0.2}$Cu$_{0.2}$Zn$_{0.2}]$Cr$_2$O$_4$ (MMgNCZCr$_2$O$_4$),  [Mn$_{0.2}$Co$_{0.2}$Mg$_{0.2}$Cu$_{0.2}$Zn$_{0.2}]$Cr$_2$O$_4$ (MCMgCZCr$_2$O$_4$),  [Mn$_{0.2}$Co$_{0.2}$Ni$_{0.2}$Mg$_{0.2}$Zn$_{0.2}]$Cr$_2$O$_4$ (MCNMgZCr$_2$O$_4$), and [Mn$_{0.2}$Co$_{0.2}$Ni$_{0.2}$Cu$_{0.2}$Mg$_{0.2}]$Cr$_2$O$_4$ (MCNCMgCr$_2$O$_4$). These compounds have the general formula $A^5$Cr$_2$O$_4$, where $A^5$ denotes an equiatomic mixture of five distinct divalent transition metals-chosen from Mg, Mn, Co, Ni, Cu, and Zn-occupying the tetrahedral $A$-site. The compounds were prepared via conventional solid-state reaction (see Methods for synthesis details).
Room-temperature powder X-ray diffraction (XRD) measurements, followed by Rietveld refinement using the FullProf program~\cite{Carvajal:2001p12}, confirm that all compounds crystallize in the normal cubic spinel structure with space group $Fd\bar{3}m$. The corresponding diffraction patterns with refinement are presented in Extended Data Figure 1. Energy dispersive X-ray spectroscopy (EDS) mapping using scanning electron microscopy (SEM) and scanning transmission electron microscopy (STEM) of the MCNCZCr$_2$O$_4$ sample confirms that all $A$-site cations are homogeneously distributed over both micrometer and nanometer length scales, respectively and there is no chemical phase separation at room temperature. Soft X-ray absorption spectroscopy of the transition metal $L_{3,2}$ edges [Extended Data Figure 1] for MCNCZCr$_2$O$_4$ and MCNCMgCr$_2$O$_4$ further confirmed that Cr$^{3+}$ ions occupy the octahedral \textit{B}-sites and the five divalent cations are distributed across the tetrahedral $A$-sites, further testifying the normal spinel structure~\cite{Nevgi:2025p2041}. 

{\color{blue}\bf Observation of structural transitions:} To elucidate the structural evolution below room temperature, we focus on the representative composition MCNCZCr$_2$O$_4$ (Mn, Co, Ni, Cu, Zn at the $A$-site), examining its temperature-dependent behavior. Upon cooling, XRD measurements reveal two successive structural phase transitions: the first near 100 K and the second around 40 K [{\bf Fig}.\ref{fig: Fig 2}a, b]. A key signature of these transitions is the evolution of the diffraction peak centered at $\sim$57$^\circ$, which in the cubic phase arises from overlapping reflections of the (511) and (333) planes. Below 100 K, this peak begins to split, with increasing separation upon further cooling, indicative of symmetry lowering. Heat capacity measurements, discussed in the later part of the manuscript, further corroborate these structural transitions.

\begin{figure*}
\begin{center}
\includegraphics[width=0.9\textwidth ]{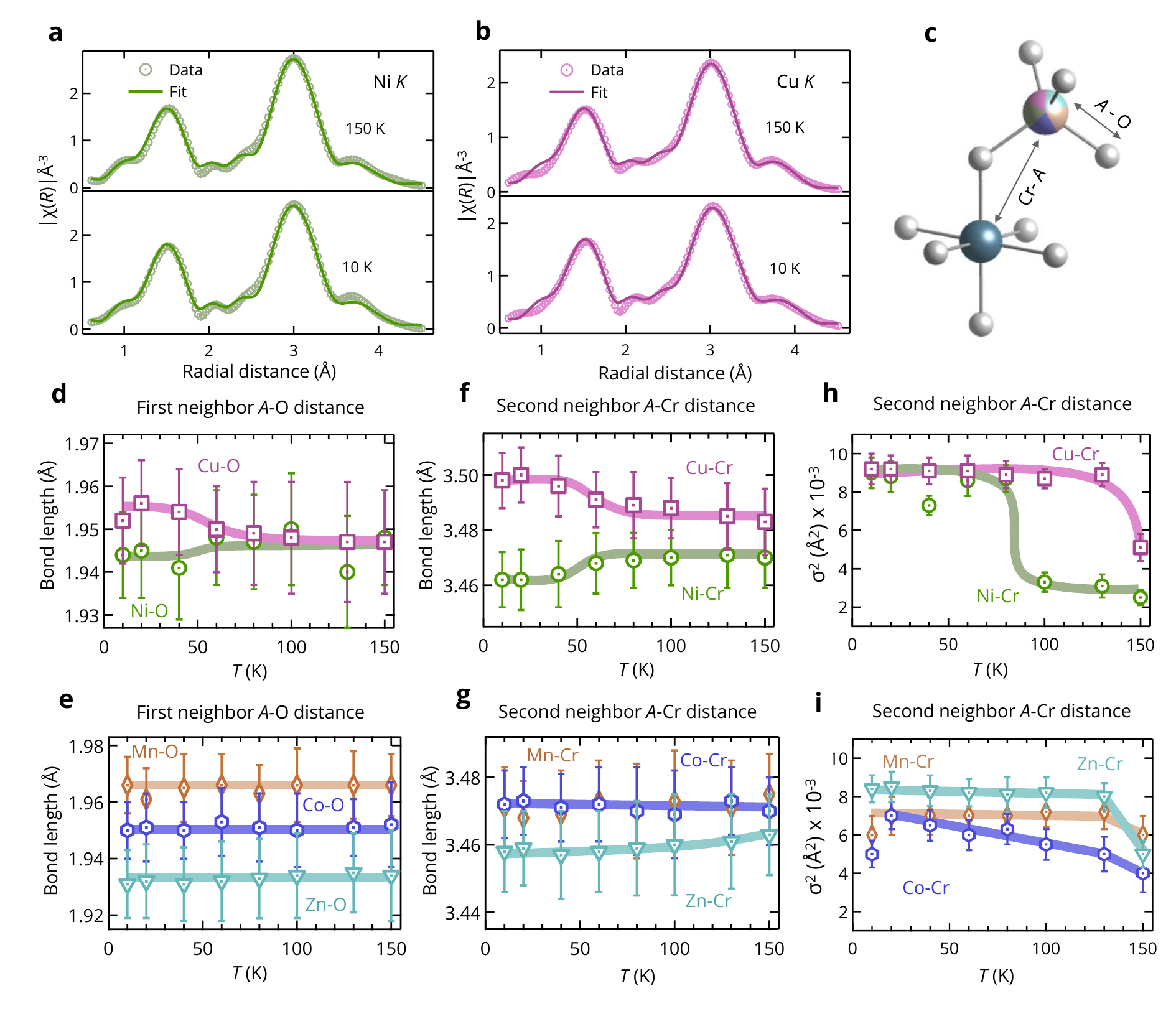}
\caption{\textbf{Temperature-dependent variation of local structural distortions probed by EXAFS.}
 Fourier transformed EXAFS spectra along with fittings at 150 K and 10 K, obtained at the  \textbf{a} Ni \textit{K} and  \textbf{b} Cu \textit{K} edges. \textbf{c}, Schematic describing the first neighbor \textit{A}-O and the second neighbor \textit{A}-Cr bond length from \textit{A} site [\textit{A} = Mn, Co, Ni, Cu, and Zn]. The variation of bond lengths in the first coordination shell of \textbf{d}, J-T active Ni and Cu and \textbf{e}, non-J-T active Mn, Co, and Zn cations. The variation of bond lengths in the second coordination shell of \textbf{f}, J-T active Ni and Cu and \textbf{g}, non-J-T active Mn, Co, and Zn cations. Variation of mean squared displacement $\sigma^{2}$ in the second neighbor (\textit{A}-Cr) bond distances in \textbf{h} J-T active Ni and Cu and \textbf{i} non-J-T active Mn, Co, and Zn cations. Additional data have been plotted in Extended Data Figure 2.}
\label{Fig3}
\end{center}
\end{figure*}

To resolve the nature of the low-temperature phases, we performed detailed peak fitting of the $\sim$57$^\circ$ region using models corresponding to cubic, tetragonal, and orthorhombic symmetries. In the cubic phase, the interplanar spacing for the (511) and (333) planes is the same, producing a single peak. Transition to tetragonal symmetry lifts this degeneracy, yielding three distinct reflections: (511), (115), and (333). Further reduction to orthorhombic symmetry results in four separate peaks: (511), (151), (115), and (333). Above 100 K, a single-peak fit confirms cubic symmetry. Between 100 K and 40 K, three peaks are required, consistent with a tetragonal phase. Below 40 K, four peaks emerge, signaling the onset of orthorhombic distortion. Rietveld refinements [{\bf Fig}.\ref{fig: Fig 2}b] corroborate this progression: at 300 K, the structure is purely cubic; at 60 K, a phase mixture comprising 21.44\% cubic and 78.56\% tetragonal ($I4_1/amd$) components is observed. We attempted the refinement of the 20 K pattern using different combinations of cubic, tetragonal, and orthorhombic phases. We find that the best refinement is obtained by a combination of 25.85\% tetragonal and 74.15\% orthorhombic ($Fddd$) phases. The scenario of chemical phase segregation leading to the observation of coexistence of these two structural phases can be ruled out, as it would require long-range atomic diffusion. Since the structure is homogeneous at room temperature, lowering the temperature would not provide the necessary diffusion energy, making such processes highly improbable at low temperatures. Moreover, high entropy materials are well known for the ``sluggish diffusion” effect, as severe varying lattice distortion leads to high diffusion barriers even at higher temperatures~\cite{Murty:2019book}. Overall, the observation of structural phase coexistence at lower temperatures indicates that the transitions proceed in a continuous and gradual manner, rather than an abrupt one.

 We now turn our attention to the mechanism underlying the cubic-to-tetragonal phase transition observed at 100 K. As summarized in {\bf Fig}.\ref{Fig1}d, the parent member ZnCr$_2$O$_4$ undergoes a spin-Peierls transition around 12 K, resulting in a coexistence of tetragonal and orthorhombic phases~\cite{Jaubert:2025p086702}. In contrast, parent compounds containing Jahn-Teller active Ni$^{2+}$ and Cu$^{2+}$ ions exhibit a purely orbital-ordering-driven transition from cubic to tetragonal symmetry, without accompanying spin ordering~\cite{Kennedy:2008p2227,Suchomel:2012p054406}. 
 Magnetic measurements (discussed later in this paper) reveal a transition to a ferrimagnetic state near 40 K in present case, suggesting that orbital interactions, rather than spin correlations, primarily drive the structural transition observed at 100 K.
It is noteworthy that orbital ordering in solids does not necessarily require direct connectivity between J-T ions via shared oxygen ions. As emphasized by Khomskii in his seminal textbook~\cite{Khomskii:2014book}, local distortions can generate long-range strain fields that influence the orbital occupation of distant ions, thereby promoting cooperative orbital ordering.
Interestingly, the nature of the tetragonal distortion differs between CuCr$_2$O$_4$ and NiCr$_2$O$_4$, reflecting the distinct electronic configurations of Cu$^{2+}$ ($d^9$) and Ni$^{2+}$ ($d^8$). In CuCr$_2$O$_4$, the tetragonal distortion is characterized by a ratio $c/a^*<$ 1, where $a^*=\sqrt2 a$, effectively lifting the degeneracy of the $d$-orbitals~\cite{Kennedy:2008p2227}. In contrast, NiCr$_2$O$_4$ requires an elongation along the $c$-axis ($c/a^*>$ 1) to lift the orbital degeneracy associated with the $d^8$ configuration, as shown in {\bf Fig}.\ref{fig: Fig 2}c~\cite{Suchomel:2012p054406}. Rietveld refinement of the crystal structure at 60 K yields lattice parameters of 
$c$ = 8.27 \AA\ and $a$ = 5.92 \AA\,  confirming a tetragonal compression consistent with CuCr$_2$O$_4$-like behavior. This raises an intriguing question: how is the orbital degeneracy of Ni$^{2+}$ lifted in the absence of the NiCr$_2$O$_4$ like elongation? Given that the remaining $A$-site cations possess either fully filled or half-filled $d$-orbitals, it is plausible that elemental-dependent local distortions vary differently across the structural transitions~\cite{Nevgi:2025p2041}. To probe these element-specific distortions, we employ element-specific extended X-ray absorption fine structure (EXAFS) spectroscopy~\cite{Rehr:2000p621}, which offers sensitivity to local atomic environments and enables disentangling the contributions of individual cations to the overall structural transition.

\begin{figure*}
\begin{center}
\includegraphics[width=\textwidth ]{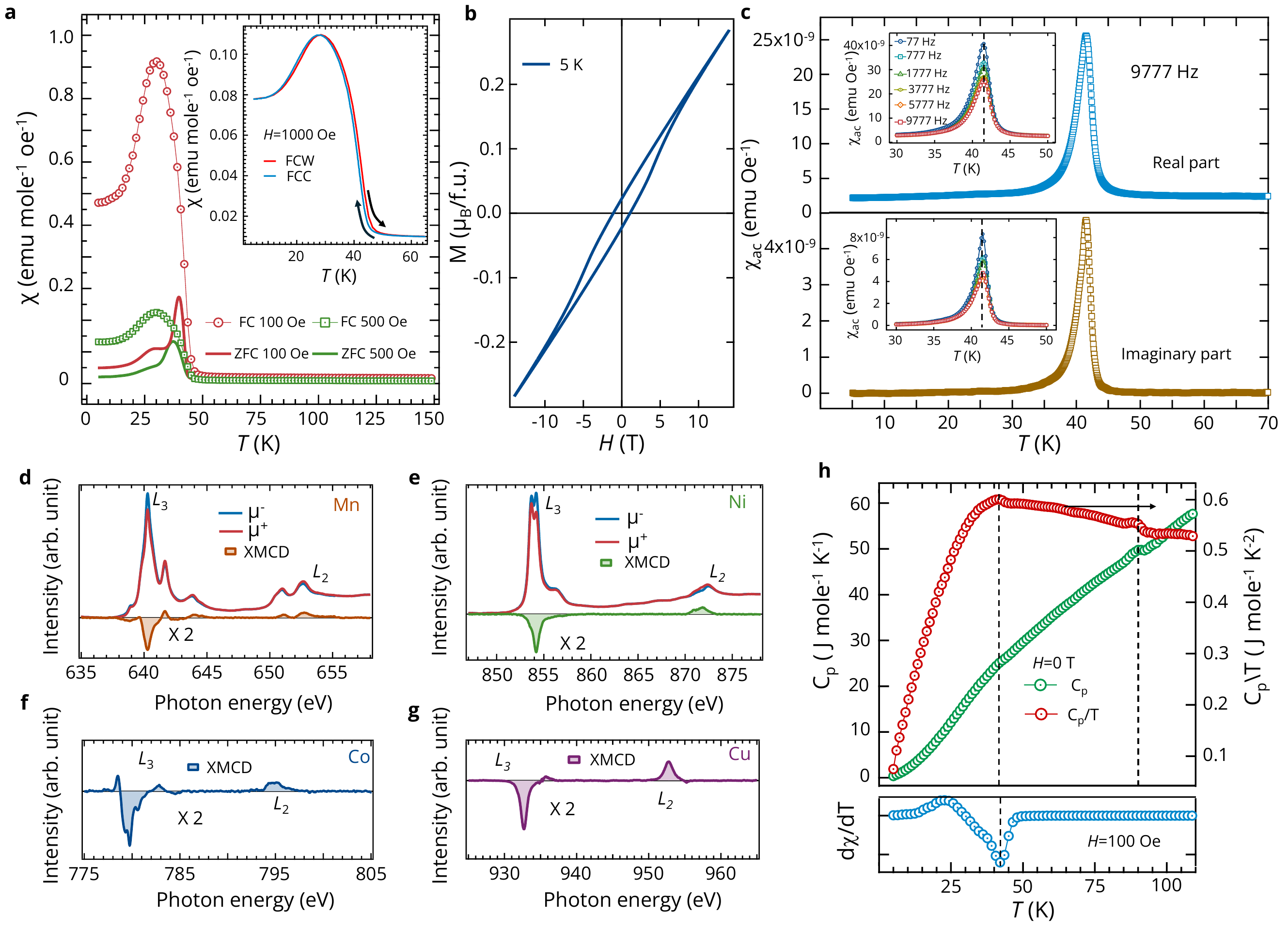}
\caption{\textbf{Magnetic and heat capacity measurements of MCNCZCr$_2$O$_4$.} \textbf{a}, Temperature-dependent magnetic susceptibility at 100 Oe and 500 Oe measured in ZFC and FC cycles. Inset shows magnetic susceptibility at 1000 Oe in field-cooled cooling (FCC) and field-cooled warming (FCW) protocols. The hysteresis between FCC and FCW indicating magneto-structural transition. \textbf{b}, Field-dependent ZFC magnetization at 5 K measured upto $\pm$14 T magnetic field. \textbf{c}, Temperature-dependent AC susceptibility at 9777 Hz frequency showing characteristics peak around 40 K for both the real and imaginary part. Inset shows AC susceptibility at various frequency for both real and imaginary part around the peak temperature. XAS spectra in right circularly ($\mu^{+}$) and left circularly ($\mu^{-}$) polarized light and XMCD signal for \textbf{d} Mn $L_{3,2}$, \textbf{e} Ni $L_{3,2}$.
XMCD for \textbf{f}, Co $L_{3,2}$, and \textbf{g}, Cu $L_{3,2}$ edges. All XMCD measurements have been carried out at 2 K in presence of 6 T field.  \textbf{h}, Temperature dependent heat capacity ($C_P$ left axis of upper panel) and $C_P/T$ (right axis of upper panel) measured under zero field shows two features around 90 K and 40 K. The lower panel shows the first-derivative of susceptibility measured under 100 Oe. The dip indicates magnetic transition temperature.}
\label{Fig4}
\end{center}
\end{figure*}

 {\color{blue}\bf Probing the local structure in MCNCZCr$_2$O$_4$:}
 To probe local structural evolution, we performed EXAFS measurements (experimental details are in the Method section) at the $K$-edges of Mn, Co, Ni, Cu, Zn, and Cr across 150–10 K, spanning the temperature range where XRD identified structural transitions. Representative Fourier transforms for Cu and Ni are shown in {\bf Fig}.\ref{Fig3}a,b; remaining data are shown in Extended Data Figure 2. 
 The first and second peaks in the $A$ $K$-edge spectra correspond to $A$-O and $A$-Cr distances, respectively [see {\bf Fig}.\ref{Fig3}c].  In the cubic spinel $A$Cr$_2$O$_4$, $A$-site cations are tetrahedrally coordinated by four O and find twelve Cr atoms as second neighbors. In lower-symmetry tetragonal or orthorhombic phases, the twelve $A$-Cr distances split into 8 and 4 coordination numbers. 
 Although XRD reveals phase coexistence cubic/tetragonal below 100 K and tetragonal/orthorhombic below 40 K in MCNCZCr$_2$O$_4$, the EXAFS spectra [Fig.\ref{Fig3}(a),(b)] show small changes across these transitions. This indicates that the symmetry-lowering distortions fall below the EXAFS resolution limit [$\Delta R$ = $\pi/2\Delta k$ = 0.16 \AA; see Methods section]. To maintain consistency and prevent over-parameterization of the EXAFS fitting, all spectra were therefore modeled using the highest-symmetry cubic structure. As a result, the extracted bond lengths should be interpreted as averaged values. In such scenarios, the variation in the Debye-Waller factor ($\sigma^2$) is used as an indicator of the local structural distortions~\cite{Booth:1998p853,Downward:2005p106401}. Similarly, we have used a cubic model to fit the Cr $K$ edge at all temperatures as the symmetry lowering distortions are below the EXAFS resolution limit of 0.2 \AA \ (see Methods section).

 {\bf Fig}.\ref{Fig3}d,e present the first-neighbor \textit{A}-O bond distances obtained from EXAFS fitting for J-T active and inactive cations, respectively. These distances exhibit a pronounced dependence on the identity of the $A$-site ion, even at a fixed temperature. Most notably, the Ni-O and Cu-O bond lengths show subtle but distinctly opposing deviations below 100 K, which is more pronounced below 50 K, coinciding structural transitions observed in X-ray diffraction. A more pronounced antagonistic behavior is observed in the second-neighbor Ni-Cr and Cu-Cr bond distances [{\bf Fig}.\ref{Fig3}f]. 
  The signature of cubic-to-tetragonal phase transition is even more clearly manifested in the temperature dependence of the $\sigma^2$ for the Ni-Cr distance [{\bf Fig}.\ref{Fig3}h]. For Cu, $\sigma^2$ begins to rise below 130 K, indicating the onset of local J-T distortions slightly preceding the global symmetry breaking. In contrast, the local coordination environments around Mn, Co remain remarkably unchanged across the entire temperature range in both coordination shells, as shown in {\bf Fig}.\ref{Fig3}e,g,i. For the Zn,  all bond distances and $\sigma^2$ remains constant below 130 K. For the Cr absorber [Extended Data Figure 2], we find a slight increase of $\sigma^2$ for the Cr-O bond below 50 K, concinding with the 40 K structural transition. Taken together, these results demonstrate that the symmetry-lowering transition selectively induces opposing distortions in both first- and second-shell bond lengths around the J-T active cations Ni$^{2+}$ and Cu$^{2+}$, effectively lifting their orbital degeneracy, as schematically illustrated in {\bf Fig}.\ref{fig: Fig 2}c. Due to the counterbalancing nature of these distortions and the structural inertness of Mn, Co, and Zn sites, the overall changes in lattice parameters across both transitions remain minimal.

{\color{blue}\bf Simultaneous magnetic and structural transitions:} 
Following the identification of a structural symmetry-lowering transition at 100 K associated with orbital ordering, we next investigate whether the tetragonal-to-orthorhombic structural transition is concomitant with a magnetic phase transition. As summarized in {\bf Fig}.\ref{Fig1}d, both parent compounds, NiCr$_2$O$_4$ and CuCr$_2$O$_4$, exhibit a transition from a tetragonal paramagnetic phase to an orthorhombic ferrimagnetic phase upon cooling~\cite{Suchomel:2012p054406}. {\bf Fig}.\ref{Fig4}a presents field-cooled (FC) and zero-field-cooled (ZFC) magnetic susceptibility data measured under applied fields of 100 Oe and 500 Oe. A sharp increase of $\chi$ is observed at 40 K, indicative of a ferrimagnetic transition. This coincides with the emergence of the orthorhombic phase observed in XRD measurements, accompanied by pronounced alterations in the local environments of Cu and Ni as revealed by EXAFS, testifying to a strongly coupled magnetostructural transition.  
Magnetostructural transitions are typically first order in nature and are often accompanied by thermal hysteresis near the transition~\cite{roy:2013p183201,lewis:2016p323002}. Our magnetization measurements in the presence of a 1000 Oe field during cooling from 300 K to 5 K and then the warming run to room temperature (inset of Fig.~\ref{Fig4}a) also find a small hysteresis near 40 K.  This transition also shifts to a higher temperature upon application of a magnetic field, further confirming the presence of a magnetostrutcural coupling. 

We also observe a subtle anomaly in the ZFC curve at 28 K in 100 Oe and 500 Oe, suggesting the emergence of non-collinear spin configurations, likely due to spin canting, as reported in other parent $A$Cr$_2$O$_4$ systems ($A$ = Mn, Co)~\cite{Yamasaki:2006p207204,Mufti:2010p075902}. We also analyzed the susceptibility measured under a 5000 Oe field using molecular field theory for ferrimagnets, which confirms that all magnetic ions (Cr$^{3+}$, Mn$^{2+}$,  Co$^{2+}$,  Ni$^{2+}$,  Cu$^{2+}$) contribute to the observed magnetism~\cite{Nevgi:2025p2041}. 
To further elucidate the nature of the magnetic transition, we performed isothermal magnetization ($M$-$H$) measurements at several temperatures. As anticipated, the $M$-$H$ curve recorded at 55 K-above the transition temperature-exhibits no hysteresis [Extended Data Figure 3], confirming the paramagnetic state. In contrast, measurements at 32 K and 20 K reveal characteristic hysteresis loops indicative of ferrimagnetic ordering.  Notably, the magnetization ($M$-$H$) loop at 5 K measured up to $\pm$14 T does not exhibit saturation {\bf Fig}.\ref{Fig4}b), and a pronounced irreversibility between the forward and reverse branches persists up to $\pm$10 T [inset of {\bf Fig}.\ref{Fig4}a]. Such behavior has previously been attributed to several factors, such as competing antiferromagnetic interactions within a non-collinear ferrimagnetic phase~\cite{Sharma:2023p161901}, magnetic inhomogeneity~\cite{Musico:2019p104416}, or spin-glass-like dynamics~\cite{Sampathkumaran:2002p180401}. Recent reports have also identified spin-glass transitions in MnCr$_2$O$_4$ and CoCr$_2$O$_4$ just above their ferrimagnetic ordering temperatures~\cite{das:2023pL100414}. Motivated by these observations, we performed AC susceptibility measurements across a broad frequency range (77 Hz to 9777 Hz). A sharp peak near 41 K is evident in both the real and imaginary components ({\bf Fig}.\ref{Fig4}c), with no discernible anomalies at lower temperatures. Crucially, the peak position remains invariant with frequency [inset of Fig.\ref{Fig4}c, and Extended Data Figure 3], thereby ruling out spin-glass behavior and further establishing the appearance of a long-range magnetic ordering.

 The spin configurations within the FIM phase in the presence of both Ni and Cu are expected to be highly complex~\cite{reehuis:2015p024407,Reehuis:2024p054414}. While neutron diffraction under applied magnetic fields would be essential to fully resolve these arrangements, we probed the relative alignment of the $A$-site cations (parallel vs. antiparallel) using element-specific X-ray magnetic circular dichroism (XMCD) measurements at 2 K under a 6 T magnetic field. The XMCD signal, defined as the difference between absorption spectra for right- and left-circularly polarized X-rays ($\mu^+ - \mu^-$; see {\bf Fig}.\ref{Fig4}c-d), is directly proportional to the net magnetization of the probed element~\cite{Stohr:2006}. As shown in {\bf Fig}.\ref{Fig4}c–f, the leading edges of the XMCD spectra for Mn, Co, Ni, and Cu exhibit the same sign (negative at the $L_3$ edge and positive at the $L_2$ edge), indicating that all $A$-site spins are aligned parallel to each other and to the direction of the applied field. In contrast, no reliable XMCD signal is observed for the Cr $L_{3,2}$ edges under identical conditions (data not shown), suggesting strong antiferromagnetic coupling among Cr ions, likely mediated by direct exchange interactions~\cite{reehuis:2015p024407}.

The simultaneous structural and magnetic transitions in parent compounds NiCr$_2$O$_4$ and CuCr$_2$O$_4$ result in a small feature at the transition temperature in heat capacity measurement~\cite{Suchomel:2012p054406}. Notably, heat capacity ($C_P$) very often does not show any features at the magnetic transitions in CCOs, as the entropy is released over a broad range of temperatures~\cite{Zhang:2019p3705}. Our zero-field $C_P$ data (Fig. \ref{Fig4}h) and $C_P/T$ versus $T$ plot [right axis of Fig. \ref{Fig4}h] reveal two characteristic features  at 41.8 K and 90 K. The 41.8 K feature coincides with the simultaneous paramagnetic-to-ferrimagnetic and tetragonal-to-orthorhombic transitions, as observed in diffraction and magnetic measurements. The cubic-to-tetragonal structural transition gives rise to a small peak in $C_P$ at 90 K, which remains insensitive to the application of a 9 T field, as expected for a pure structural transition. Most importantly, the 41.8 K feature shifts to 48.4 K in the presence of a 9 T field, which is also very close  to  the transition temperature found from the magnetization measurement under 7 T field. Overall, these physical property measurements along with structural probes, directly demonstrate the observation of a coupled magnetostructural transition under a compositionally complex setting.

\begin{figure*}
\begin{center}
\includegraphics[width=\textwidth ]{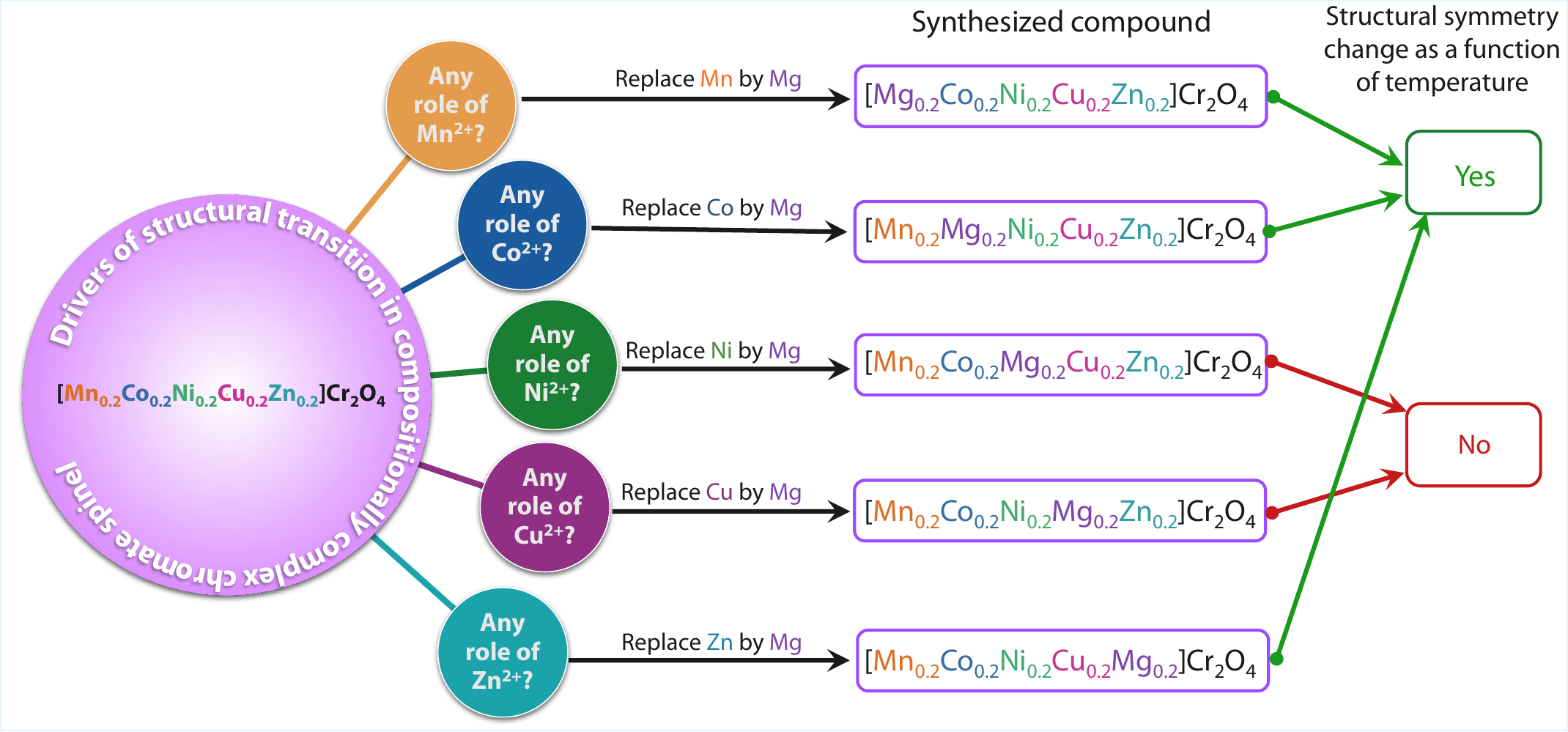}
\caption{{\bf Drivers of structural transitions}. A schematic to demonstrate the substitutional strategy to synthesize a series of compositionally complex chromate spinels. The presence/absence of temperature-driven structural transitions has been also mentioned. Temperature-dependent XRD data has been shown in Extended Data Figure 4.}
\label{Fig5}
\end{center}
\end{figure*}

{\color{blue}\bf Discussions:} Having established that MCNCZCr$_2$O$_4$ undergoes coupled phase transitions at approximately 100 K and 40 K, we next investigate the role of individual $A$-site cations in driving these structural transitions. Understanding these contributions is essential for developing a generalizable strategy for designing HEOs with targeted properties. To this end, we employed a systematic substitutional approach in which each magnetic and/or (J-T) active $A$-site cation in MCNCZCr$_2$O$_4$ was selectively replaced with non-magnetic, non J-T active Mg$^{2+}$ ({\bf Fig}.\ref{Fig5}). This strategy enables isolation of the influence of specific cations on the transition behavior.

As shown in Extended Data Figure 1, all of these compounds crystallize in the cubic phase at room temperature. Temperature-dependent XRD measurements were performed down to 20 K for each substituted compound. The evolution of the diffraction patterns is presented in Extended Data Figure 4, and the presence or absence of structural transitions is summarized in {\bf Fig}.\ref{Fig5}. 
 A consistent and compelling trend emerges: compositions containing either Ni$^{2+}$([Mn$_{0.2}$Co$_{0.2}$Ni$_{0.2}$Mg$_{0.2}$Zn$_{0.2}$]Cr$_2$O$_4$) or Cu$^{2+}$ ([Mn$_{0.2}$Co$_{0.2}$Mg$_{0.2}$Cu$_{0.2}$Zn$_{0.2}$]Cr$_2$O$_4$) alone retain the cubic structure, showing no evidence of structural transitions down to 20 K. In contrast, compositions containing both Ni$^{2+}$ and Cu$^{2+}$ exhibit temperature-driven structural phase transitions that closely mirror the behavior observed in MCNCZCr$_2$O$_4$ [Extended Data Figure 4]. Remarkably, the onset temperatures of these transitions are very similar across all substituted compositions (onset $\approx$ 100 K). These findings demonstrate that dual J-T activity is a necessary condition for inducing the observed structural transitions in this class of materials.

We also anticipate the opposing local distortions around Ni and Cu lead to the structural phase coexistence, instead of a 100\% phase transformation. To test for this, we have further synthesized  [Mn$_{0.2}$Co$_{0.2}$Cu$_{0.4}$Zn$_{0.2}$]Cr$_{2}$O$_{4}$ by substituting Ni by Cu in [Mn$_{0.2}$Co$_{0.2}$Ni$_{0.2}$Cu$_{0.2}$Zn$_{0.2}$]Cr$_{2}$O$_{4}$. The XRD pattern recorded at 300 K confirmed that the compound remains cubic. Most importantly, the compound exhibits a 100\% orthorhombic phase at 20 K, even though the configurational entropy of this composition is below the conventional threshold for high-entropy compounds ($S_\mathrm{config} \geq$ 1.5$R$, $R$ is the universal gas constant).
On the other hand, structural phase transitions are absent in the compositions [Mn$_{0.2}$Co$_{0.2}$Mg$_{0.2}$Cu$_{0.2}$Zn$_{0.2}$]Cr$_{2}$O$_{4}$ and [Mn$_{0.2}$Co$_{0.2}$Mg$_{0.2}$Ni$_{0.2}$Zn$_{0.2}$]Cr$_{2}$O$_{4}$, which have configurational entropy above the threshold ($S_\mathrm{config}$ = 1.61R). Thus, the temperature-driven structural phase coexistence in compositions with $S_\mathrm{config}$ = 1.61$R$ is governed by the opposing Jahn-Teller active Ni$^{2+}$ and Cu$^{2+}$.

  To summarize, this work provides direct experimental evidence for temperature-induced coupled phase transitions in compositionally complex chromate spinels, uniquely driven by the cooperative interplay of two J-T active ions, Ni$^{2+}$ and Cu$^{2+}$. Element-specific EXAFS measurements reveal opposing, temperature-dependent bond length modulations around these ions across these transitions, establishing a rare case of cooperative transitions emerging from local competition in a chemically disordered crystalline matrix, which we term as `cooperation through competition'. Beyond demonstrating a robust design strategy for tuning structural responses in complex oxides, this work advances the fundamental understanding of symmetry breaking in high-entropy systems and opens new avenues for engineering functional materials through targeted cation chemistry in a high entropy setting.

\section{Methods}
 The polycrystalline compositionally complex chromate spinel compositions were synthesized by the conventional solid-state reaction starting with stoichiometric amounts of Cr$_2$O$_3$ and respective \textit{A}O oxides. The first annealing was carried out at 900$^\circ$ C, for 12 hours, followed by a second heat treatment at 1300$^\circ$ C, again for 12 hours with intermediate grindings. The phase purity of all the samples were checked by powder XRD using a laboratory-based Rigaku Smartlab diffractometer. Microstructural analysis and elemental mapping were performed using field emission scanning electron microscopy (FE-SEM) and scanning transmission electron microscopy (STEM) in micrometer and nanometer length scale equipped with energy dispersive X-ray (EDX).Temperature-dependent XRD measurements were carried out between 20 K to 300 K temperature range using a laboratory-based Bruker D8 Discover diffractometer and also in the Indian beamline (BL-18B) at the Photon Factory, KEK, Japan. DC and AC magnetic measurements have been carried out using a SQUID magnetometer [Quantum Design] and a Physical Property Measurement System  [Quantum Design]. Heat capacity measurements were performed using a Physical Property Measurement System [Quantum Design]. X-ray absorption spectroscopy (XAS) spectra for $L_{3,2}$ edges of Cr, Mn, Co, Ni, Cu were recorded in total electron yield mode at the beamline BL 4.0.2 of the
Advanced Light Source, USA. XMCD at the $L_{3,2}$ edges of Cr, Mn, Co, Ni, and Cu was carried out at BL29-BOREAS beamline, ALBA synchrotron light source, Spain, at 2 K under a 6 T magnetic field.

EXAFS measurements were performed at Cr, Mn, Co, Ni, Cu and Zn \textit{K}-edges at temperatures between 150 K and 10 K at the P65 beamline of the PETRA III Synchrotron Source (DESY, Hamburg, Germany). The fine powder was uniformly spread onto scotch tape, and absorbers were prepared by stacking multiple layers as required. Gas ionization chambers were employed as detectors to record incident and transmitted photon intensities. To reduce statistical noise, each edge was scanned at least three times. The software DEMETER Suite was used to carry out the analysis~\cite{Ravel:2005p537}.

\textbf{EXAFS analysis:} The EXAFS spectra were calibrated using standard metal foils. To avoid interference from nearby absorption edges, we truncated the spectra in an appropriate range. For the \textit{A} $K$-edge spectra (where \textit{A} = Mn, Co, Ni, Cu, and Zn), we used a \textit{k} range of 2 to 12 \AA$^{-1}$ and an \textit{R} range of 1 to 4 \AA\ for the fittings. For the Cr $K$-edge, the \textit{k} range was set from 2 to 10 \AA$^{-1}$ and the \textit{R} range from 1 to 3.5 \AA \ for the fittings.

In our fitting process, we primarily considered the following parameters: coordination number (N), amplitude reduction factor ($S_0^2$), energy shift ($E_0$), change in bond length ($\Delta R$), and mean squared displacement ($\sigma^2$), also referred to as the Debye-Waller factor. Since MCNCZCr$_2$O$_4$ exhibits a normal spinel structure, we assigned all \textit{A} cations (Mn, Co, Ni, Cu, and Zn) to the tetrahedral site and Cr at the octahedral site. The intricate details of the room temperature EXAFS analysis  MCNCZCr$_2$O$_4$  have been described in our recent publication~\cite{Nevgi:2025p2041}. We have followed similar process for the EXAFS analysis of low temperatures as well. For the sake of brevity, we briefly discuss it here.

The EXAFS fittings for every metal $K$ edge were performed independently [we do not find any significant variations in bond lengths obtained through multiedge analysis~\cite{Nevgi:2025p2041}]. 
The splitting in bond lengths below 100 K was smaller than the resolution of EXAFS $\Delta R$ described as $\Delta R$ = $\pi/2\Delta k$, $\Delta k = k_{max}-k_{min}$. For all the \textit{A K} edges, this was 0.16 \AA \ ($\Delta k = k_{max}-k_{min}$ = 12-2 \AA$^{-1}$), while for Cr \textit{K} edges it was 0.2 \AA \ ($\Delta k = k_{max}-k_{min}$ = 10-2 \AA$^{-1}$). This prevented us from introducing a lower symmetry model at low temperatures. Hence, we adopted cubic spinel model for the fittings at each metal \textit{K}-edge at all temperatures down to 10 K.   During the fitting process, we used paths corresponding to direct scatterers. The values of $S_0^2$ and $E_0$ were employed from the room temperature analysis and kept constant. The values $\Delta R$ and $\sigma^2$ were varied for each path considered. The coordination number N was fixed to the stoichiometric value for every path.

\section{Description of Extended Data Figures}

\textbf{Extended Data Figure 1}: Room temperature XRD patterns [panels \textbf{a–f}] of six compositionally complex chromate spinels are shown. All of these XRD patterns have been analyzed using Rietveld refinement, which confirms that all six compositions crystallize in the normal cubic spinel phase with space group $Fd\bar3m$.

In our previous manuscript \cite{Nevgi:2025p2041}, we reported the $L_{3,2}$ edge XAS of  MCNCZCr$_{2}$O$_{4}$ at 300 K, which verified the normal spinel configuration with Mn$^{2+}$, Co$^{2+}$, Ni$^{2+}$, and Cu$^{2+}$ occupying the tetrahedral \textit{A} sites and Cr$^{3+}$ residing at the octahedral B sites. In  panels \textbf{g–k}, we compare the $L_{3,2}$ edges of the transition-metal cations in MCNCMgCr$_{2}$O$_{4}$ (Mn$_{0.2}$Co$_{0.2}$Ni$_{0.2}$Cu$_{0.2}$Mg$_{0.2}$Cr$_{2}$O$_{4}$) with those in MCNCZCr$_{2}$O$_{4}$. The spectra are very similar, confirming that MCNCMgCr$_{2}$O$_{4}$ also stabilizes in the desired normal cubic spinel structure.

\textbf{Extended Data Figure 2}: This figure serves as an extension of {\bf Fig}.\ref{Fig3} and offers further insights into the local structure of MCNCZCr$_{2}$O$_{4}$. Panels \textbf{a–d} show the magnitude of the Fourier-transformed EXAFS spectra, together with the corresponding fits, collected at 150 K and 10 K for the Mn \textit{K}, Co \textit{K}, Zn \textit{K}, and Cr \textit{K} edges, similar to the Cu \textit{K} and Ni \textit{K} edges that are presented in {\bf Fig}.\ref{Fig3}. The figure further emphasizes the local structural environment of Cr. Panels \textbf{f–i} present the temperature dependence of the first and second neighbor bond distances around Cr, along with their mean-squared displacement $\sigma^2$. Just like the Mn, Co, and Zn local environments, no significant changes in bond lengths are observed around the Cr local structure. However, the mean-squared displacement $\sigma^2$ of the first-neighbor Cr-O bond (panel g) exhibits a slight increase below 50 K, coinciding with the onset of the orthorhombic phase observed in XRD.

\textbf{Extended Data Figure 3}: This figure serves as an extension of {\bf Fig}.\ref{Fig4} and provides further information into the magnetic property measurements.
Panel \textbf{a} presents the \textit{M-H} loops at 20 K, 32 K, and 55 K. Panel \textbf{b} shows the frequency dependence of the ac susceptibility peak position (referenced in {\bf Fig}.\ref{Fig4}c), where the negligible shift with frequency rules out spin-glass freezing. Additionally, panel \textbf{c} displays the field-cooled magnetic susceptibility measured under an applied field of 100 Oe for [Mn$_{0.2}$Co$_{0.2}$Ni$_{0.2}$Cu$_{0.2}$Mg$_{0.2}$]Cr$_2$O$_4$ , which reveals a ferrimagnetic transition near 40 K, similar to that observed in MCNCZCr$_2$O$_4$.

\textbf{Extended Data Figure 4}: Temperature-dependent XRD of [Mg$_{0.2}$Co$_{0.2}$Ni$_{0.2}$Cu$_{0.2}$Zn$_{0.2}$]Cr$_2$O$_4$ and [Mn$_{0.2}$Co$_{0.2}$Ni$_{0.2}$Cu$_{0.2}$Mg$_{0.2}$]Cr$_2$O$_4$ were collected at KEK, JAPAN using a photon energy of 12 keV [$\lambda$=1.0332 \AA]. For the other three compounds, [Mn$_{0.2}$Mg$_{0.2}$Ni$_{0.2}$Cu$_{0.2}$Zn$_{0.2}$]Cr$_2$O$_4$, [Mn$_{0.2}$Co$_{0.2}$Mg$_{0.2}$Cu$_{0.2}$Zn$_{0.2}$]Cr$_2$O$_4$, and [Mn$_{0.2}$Co$_{0.2}$Ni$_{0.2}$Mg$_{0.2}$Zn$_{0.2}$]Cr$_2$O$_4$, a laboratory based diffractometer was used [$\lambda$=1.5405 \AA]. For easy comparison among these compounds, we have plotted the $x$ axis as the interplanar spacing ($d$).

Panels \textbf{a–c} display scans for the three compositions that undergo a structural transition upon cooling. The transition is evident from the splitting of the (440), (333), and (115) peaks. All three compositions contain both J-T active cations, Ni and Cu, exhibit a similar transition onset temperature between 90 K and 100 K. Panels \textbf{d–e} show scans at 300 K and 20 K for compositions containing only a single J-T active cation (either Ni or Cu), demonstrating the absence of any structural transition.

\section*{Acknowledgments}
The authors acknowledge the use of central facilities of the Department of Physics, IISc Bangalore, funded by the FIST program of DST (Department of Science and Technology), India, the central X-ray facility and advanced facility for microscopy and microanalysis (AFMM) of IISc. 
SM acknowledges funding
support from a SERB Core Research grant (Grant No. CRG/2022/001906)
and I.R.H.P.A Grant No. IPA/2020/000034.  SD, NB acknowledges funding from the Prime Minister’s Research Fellowship (PMRF), MoE, Government of India. RN thanks the Indian Institute of
Science for support through Sir C. V. Raman postdoctoral fellowship program. The authors acknowledge the use of SQUID magnetometer facility from Goa University and thank Dr. Elaine Dias for magnetic measurements. The authors also acknowledge the use of the Physical Properties Measurement  System (9 T  PPMS from M/s. Quantum Design, USA) in UGC-DAE Consortium for Scientific Research, Mumbai Centre. We acknowledge Dr. Rajeev Rawat of UGC-DAE Consortium for Scientific Research, Indore Centre for measuring the $\pm$ 14 Tesla hysteresis loop. Portions of this research were carried out at the light source PETRA III DESY, a member of the Helmholtz Association (HGF).  Financial support by the Department of Science \& Technology (Government of India) provided within the framework of the India@DESY collaboration is gratefully acknowledged.  The authors thank the Department of Science and Technology, India for the financial support and Saha Institute of Nuclear Physics and Jawaharlal Nehru Centre for Advanced Scientific Research, India, for facilitating the experiments at the Indian Beamline, Photon Factory, KEK, Japan. This research used resources of the Advanced Light Source, which is a Department
of Energy Office of Science User Facility under Contract No. DE-AC02-05CH11231.

\newpage

\renewcommand{\figurename}{Extended Data Fig}
\setcounter{figure}{0}

\begin{figure*}
\begin{center}
\includegraphics[width=0.85\textwidth ]{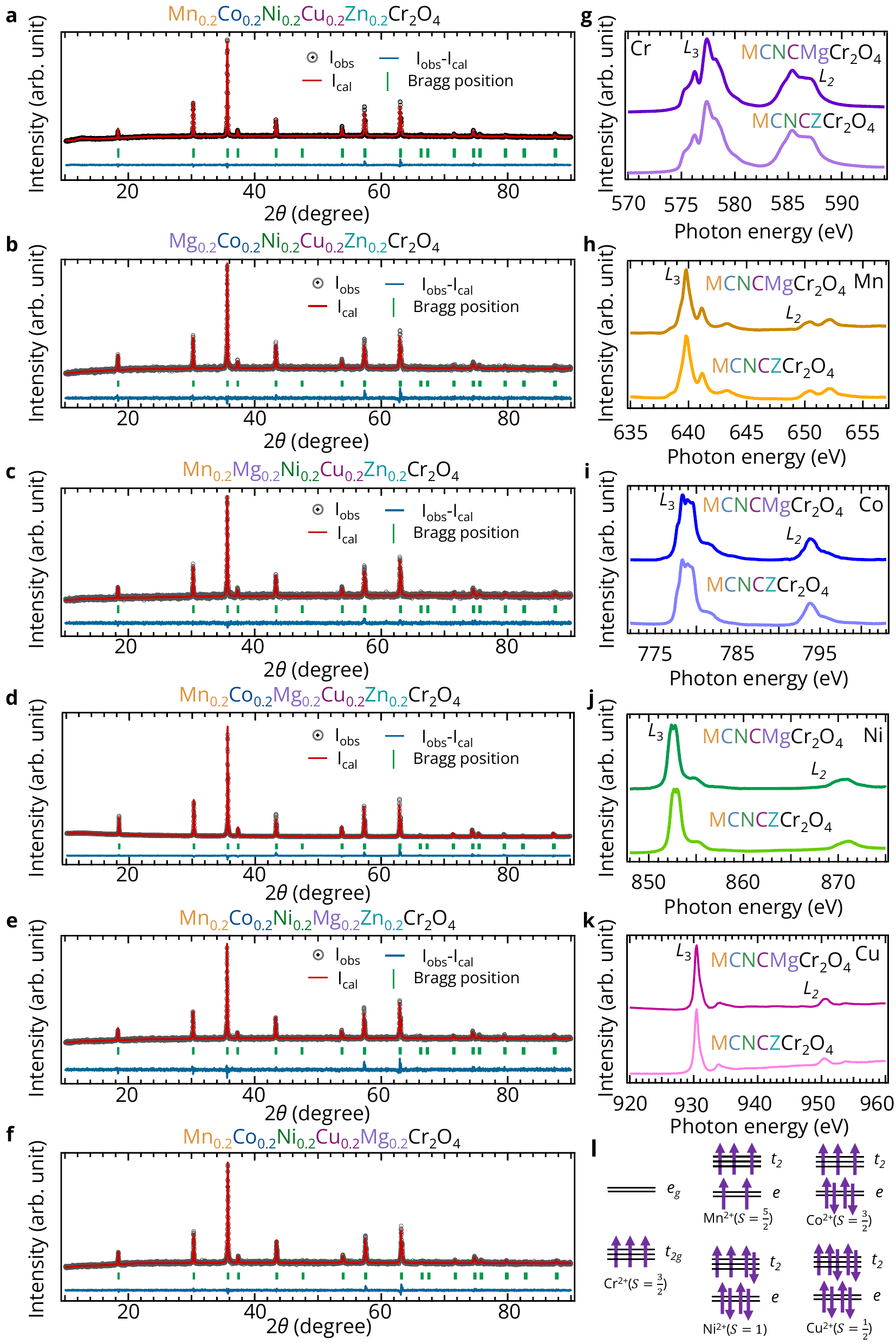}
\caption{\textbf{Structural characterization of compositionally complex chromate spinels at 300 K:} \textbf{a}-\textbf{f} Rietveld refinement of XRD patterns  
for different compositions confirming a single-phase cubic spinel structure [$Fd\bar3m$]. \textbf{g}-\textbf{k} A comparison of XAS spectra at $L_{3,2}$ edges of Cr, Mn, Co, Ni,  and Cu of Mn$_{0.2}$Co$_{0.2}$Ni$_{0.2}$Cu$_{0.2}$Mg$_{0.2}$Cr$_{2}$O$_{4}$ (MCNCMgCr$_2$O$_4$) with MCNCZCr$_2$O$_4$ showing similar spectral features at each cation. \textbf{i} Schematic representation of the expected spin configuration in octahedral (Cr$^{3+}$) and tetrahedral environments.($A^{2+}$, \textit{A} = Mn, Co, Ni, and Cu).}
\label{fig: Extended Fig 1}
\end{center}
\end{figure*}


\begin{figure*}
\begin{center}
\includegraphics[width=\textwidth ]{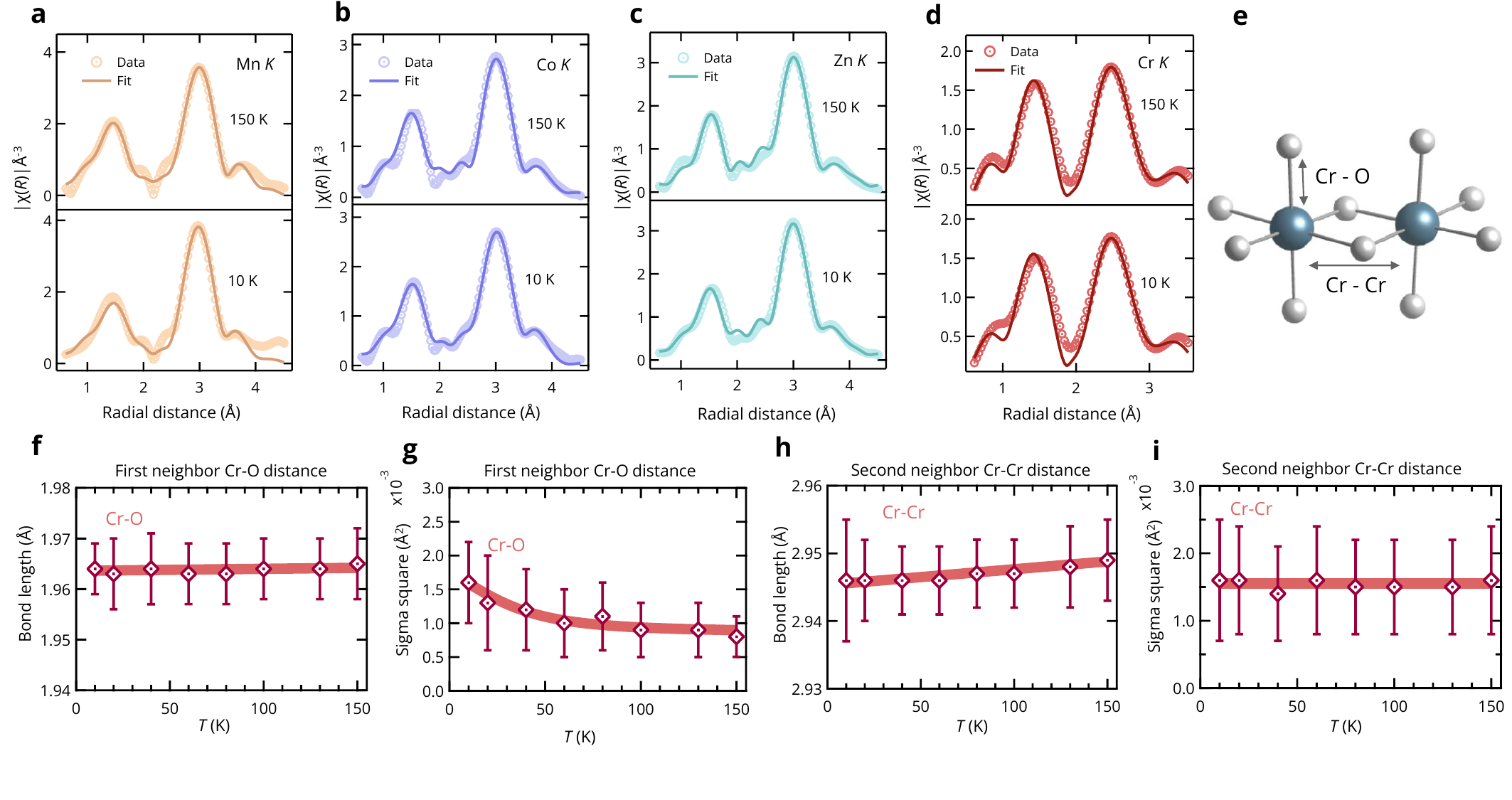}
\caption{\textbf{Local structure in MCNCZCr$_2$O$_4$:} The magnitude of the Fourier transformed EXAFS spectra along with fittings at 150
K and 10 K, obtained at the \textbf{a}-\textbf{c} \textit{A} K-edges [\textit{A} = Mn, Co, and Zn] and \textbf{d} Cr-K edge. \textbf{e}  Schematic describing the first neighbor Cr-O and the second neighbor Cr-Cr bond length from Cr site. \textbf{f}-\textbf{g} The bond length and mean squared displacement $\sigma^{2}$ variation in the first coordination shell Cr-O with temperature. \textbf{h}-\textbf{i} The bond length and mean squared displacement $\sigma^{2}$ variation in the second coordination shell Cr-Cr with temperature.}
\label{fig: Extended Fig 2}
\end{center}
\end{figure*}

\begin{figure*}
\begin{center}
\includegraphics[width=0.95\textwidth ]{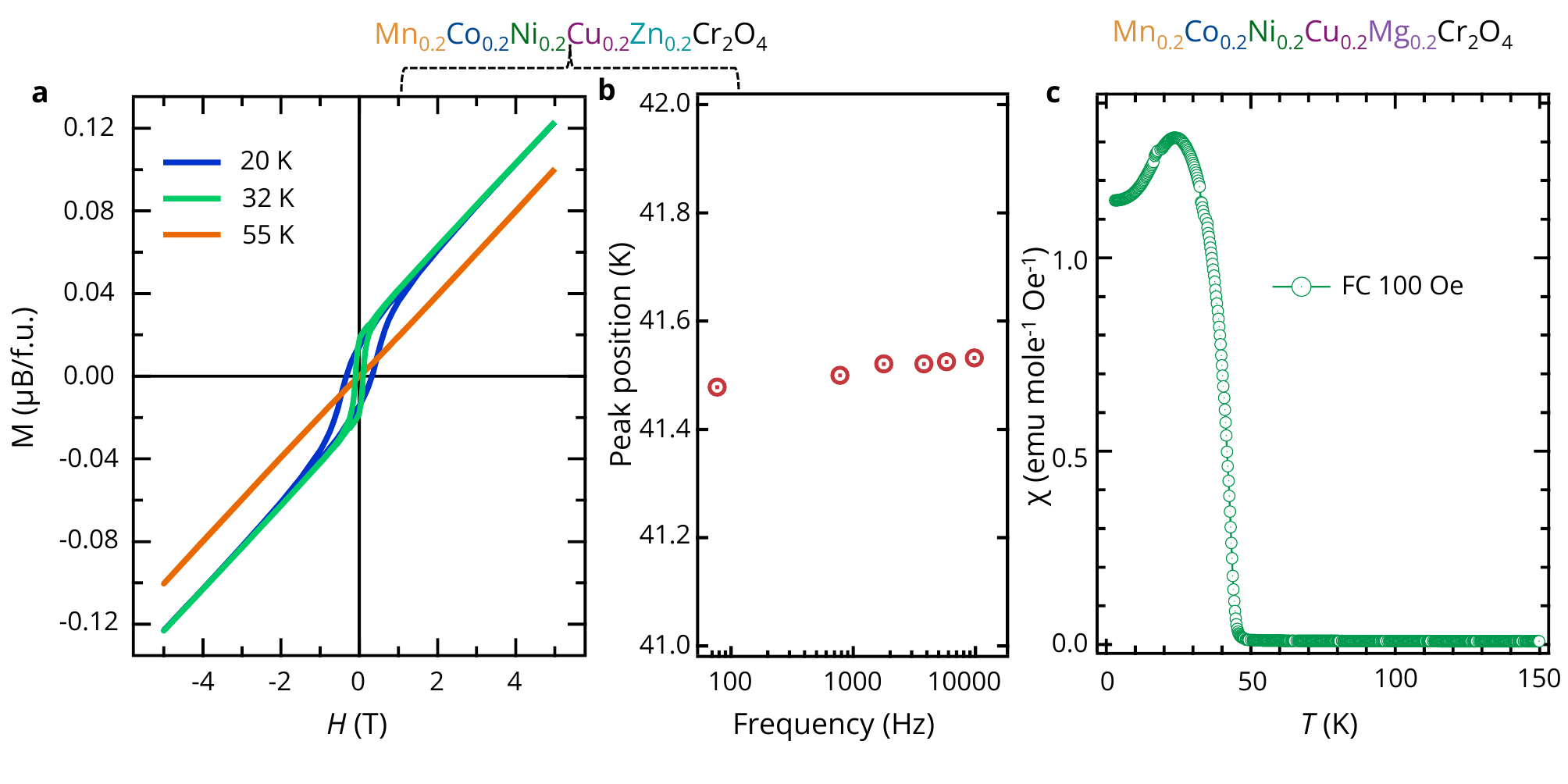}
\caption{\textbf{Magnetic property measurements:} \textbf{a} Isothermal magnetization (M–H) curves measured at 20 K, 32 K, and 55 K in ZFC, showing the evolution of magnetic response across the transition region. Hysteresis loop below 41 K indicates ferrimagnetic ordering. \textbf{b} Frequency dependence of the ac susceptibility peak position, indicating minimal shift with frequency and suggesting the absence of spin-glass freezing. \textbf{c} Field cooled magnetic susceptibility under applied field 100 Oe of Mn$_{0.2}$Co$_{0.2}$Ni$_{0.2}$Cu$_{0.2}$Mg$_{0.2}$Cr$_2$O$_4$ indicates ferrimagnetic transition near 40 K.  }
\label{fig: Extended Data Fig 3}
\end{center}
\end{figure*}


\begin{figure*}
\begin{center}
\includegraphics[width=1\textwidth ]{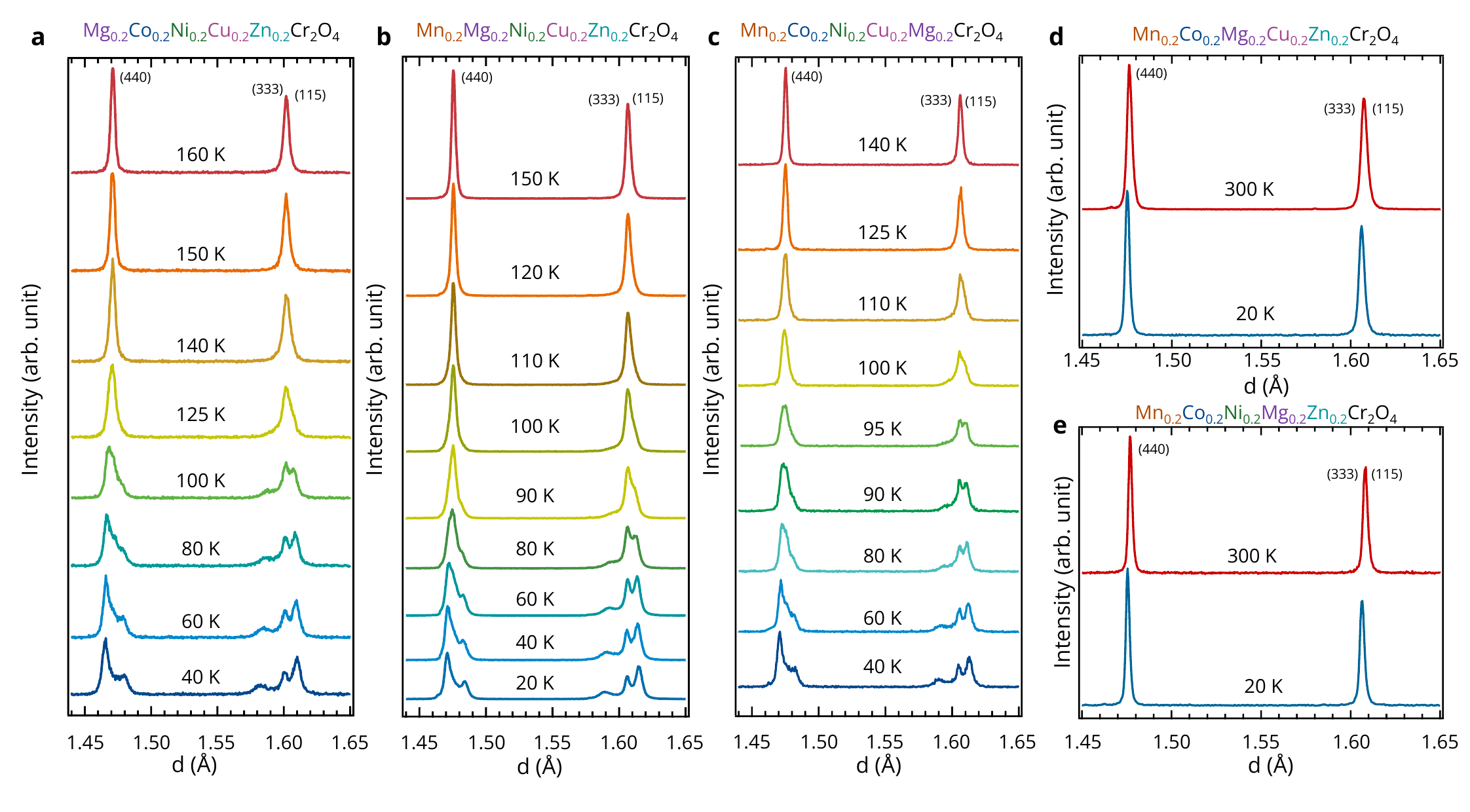}
\caption{\textbf{Temperature-dependent XRD scans of compositionally complex chromate spinels:} Selected temperature-dependent XRD patterns in the high-angle region for \textbf{a}  Mn$_{0.2}$Mg$_{0.2}$Ni$_{0.2}$Cu$_{0.2}$Zn$_{0.2}$Cr$_{2}$O$_{4}$, \textbf{b}  Mg$_{0.2}$Co$_{0.2}$Ni$_{0.2}$Cu$_{0.2}$Zn$_{0.2}$Cr$_{2}$O$_{4}$, and \textbf{c} Mn$_{0.2}$Co$_{0.2}$Ni$_{0.2}$Cu$_{0.2}$Mg$_{0.2}$Cr$_{2}$O$_{4}$. The splitting of the (440), (333) and (115) peaks below $\approx$ 100 K indicates structural phase transitions to lower symmetry phases. XRD patterns in the high-angle region collected at 300 K and 20 K for \textbf{d} Mn$_{0.2}$Co$_{0.2}$Mg$_{0.2}$Cu$_{0.2}$Zn$_{0.2}$Cr$_{2}$O$_{4}$ and \textbf{e} Mn$_{0.2}$Co$_{0.2}$Ni$_{0.2}$Mg$_{0.2}$Zn$_{0.2}$Cr$_{2}$O$_{4}$. The data confirms the cubic phase at both temperatures, indicating the absence of a structural transition in both compositions. }
\label{fig: Extended Data Fig 4}
\end{center}
\end{figure*}

\end{document}